\documentclass[a4paper,10pt]{article}
\usepackage{natbib}

\usepackage{amsmath}
\usepackage{amsfonts}
\usepackage{amssymb}
\usepackage{epsfig}

\newcommand{\real}{\rm{I\!R}}
\newcommand{\T}{^{\rm T}}

\title{Blockwise and coordinatewise thresholding to combine tests of different natures in modern ANOVA}

\author{Sylvain Sardy}
\date{}

\begin{document}

\maketitle

\begin{abstract}
We derive new tests for fixed and random ANOVA based on a thresholded point estimate.
The pivotal quantity is  the threshold that sets all the coefficients of the null hypothesis to zero.
Thresholding can be employed coordinatewise or blockwise, or both, which leads to tests with good power properties
under alternative hypotheses that are either sparse or dense.
\end{abstract}


\section{Introduction}
\label{sct:intro}

Analysis of variance (ANOVA) has something in common with thresholding regression in that
ANOVA tests a null hypothesis that some parameters are equal to zero,
and thresholding performs model selection by setting some coefficients to zero. This paper exploits this link to derive ANOVA tests based on thresholding.

While ANOVA and tests belong to the general knowledge of a statistician, thresholding is a more recent concept that we now review.
A simple way to introduce thresholding and thresholding test is to consider  the canonical regression  model
\begin{equation} \label{eq:model0}
Y_n =  \theta_n + \epsilon_n \quad  {\rm for} \quad n=1,\ldots,N,
\end{equation}
where the noise is independent standard Gaussian. 
% In that setting thresholding 
% sets some or all estimated parameters $\hat {\boldsymbol \theta}$ to zero, that is $\hat \theta_n=0$ for some or all $n \in \{1,\ldots,N\}$.
%, for those not deemed sufficiently significant.
A thresholded point estimate  of parameters ${\boldsymbol \theta}=(\theta_1,\ldots, \theta_N)$ is obtained by applying a function $\eta_\lambda$ to the data ${\bf Y}=(Y_1,\ldots,Y_N)$, that is,
\begin{equation} \label{eq:thresholded0}
\hat {\boldsymbol \theta}=\eta_\lambda({\bf Y}) 
\end{equation}
with the property that some or all entries of the  estimate are null for a large enough threshold $\lambda$.
Defining $(x)_+=\max(x,0)$, thresholding can be performed:
\begin{itemize}
 \item coordinatewise, for instance with soft-thresholding  \citep{Dono94b}: considering each $n$ in turn, estimate $\theta_n$ with
 \begin{equation}\label{eq:soft} 
%  \eta_\lambda^{\rm soft}(Y_n)=\left (1-\frac{\lambda}{|Y_n|}\right )_+ Y_n 
 \eta_\lambda^{\rm soft}(Y_n)=\left (1-\frac{\lambda}{|Y_n|}\right )_+ Y_n  = : \hat \theta_{n}.
 \end{equation}
 \item blockwise, for instance with truncated James-Stein thresholding \citep{Jame:Stei:esti:1961}: considering all entries of ${\bf Y}$ together, estimate ${\boldsymbol \theta}$ with
 \begin{equation} \label{eq:JS+}
% \eta_\lambda^{\rm JS+}({\bf Y})=\left (1-\frac{\lambda}{\|{\bf Y} \|_2^2} \right )_+ {\bf Y}
\eta_\lambda^{\rm JS+}({\bf Y})=\left (1-\frac{\lambda}{\|{\bf Y} \|_2^2} \right )_+ {\bf Y} = : \hat {\boldsymbol \theta}.
\end{equation}
%The original blockwise thresholding function can be traced back to the truncated James-Stein estimator \cite{Jame:Stei:esti:1961}  for $\lambda=N-2$. 
\end{itemize}

% 
% coordinatewise or blockwise to the vector of data by setting either some entries of $\hat {\boldsymbol \theta}$ to zero, or all of them at once. 
% \begin{eqnarray}
% \eta_\lambda^{\rm soft}(y_n)&=&\left (1-\frac{\lambda}{|y_n|}\right )_+ y_n \label{eq:soft} \\
% \eta_\lambda^{\rm hard}(y_n)&=&1_{ \{|y_n|\geq \lambda \}} y_n \label{eq:hard}
% \end{eqnarray}
% for $n=1,\ldots,N$, where $(x)_+=\max(x,0)$.
% The original blockwise thresholding function can be traced back to the truncated James-Stein estimator \cite{Jame:Stei:esti:1961} which uses
% \begin{equation} \label{eq:JS+}
% \eta_\lambda^{\rm JS+}({\bf Y})=\left (1-\frac{\lambda}{\|{\bf Y} \|_2^2} \right )_+ {\bf Y}
% \end{equation}
% for $\lambda=N-2$. %used also for wavelet smoothing \cite{Cai:adap:1999}.

The choice of the threshold $\lambda$ plays an important role in the quality of the estimation in regression
(see for instance
\citep{Dono94b}, \citep{Dono95i}, \citep{Tibs:regr:1996}, \citep{Yuan:Lin:mode:2006}, \citep{Efro:esti:2004} and \citep{ZHT07}) or to control the false discovery rate \citep{Benj:Hoch:1995}.

In this article we are interested in linear analysis of variance and testing null hypotheses
regarding factors and continuous covariates.
We derive new powerful tests based on a thresholded point estimate of the coefficients, and
we choose the threshold $\lambda_\alpha$ for the test to have the desired level $\alpha$.
Since we commented that thresholding can be performed coordinatewise or blockwise, we derive tests based on either coordinate
or  block thresholding. Or hybrids of both.
Two tests are used extensively in ANOVA: Tukey's multiple comparisons test and Fisher $F$-test.
The first one is related to coordinate thresholding and the latter to block thresholding.
One goal is to combine Tukey- and Fisher-like tests.

% 
% the choice of $\lambda_\lambda$ to control the level $\alpha$ of tests
% in an analysis of variance (ANOVA) setting.
% 
% the null hypothesis
% $$
% H_0: \ \theta_1=\ldots=\theta_N=0
% $$
%against the alternative hypothesis that at least one parameter is different from zero.

We illustrate the link between thresholding and testing on the canonical model (\ref{eq:model0}) and derive two tests for $H_0: \ \theta_1=\ldots=\theta_N=0$ using a thresholded point estimate.
%And we show how to choose their threshold to set the test to the desired level.
For simplicity we assume for now unit standard deviation for the Gaussian noise. The two tests are:
\begin{itemize}
 \item based on coordinatewise thresholding (\ref{eq:soft}): clearly, for a sample ${\bf y}$, $\lambda_{\bf y}=\max_{n=1,\ldots,N} |y_n|=\|{\bf y} \|_\infty$
 is the smallest and finite threshold that sets all the estimated parameters to zero.
 Letting $F_{\Lambda_0}$ be the distribution of that statistics under the null hypothesis, and choosing $\lambda_\alpha=F_{\Lambda_0}^{-1}(1-\alpha)$,
  the test that rejects $H_0$ if $\lambda_{\bf y}>\lambda_\alpha$ or, equivalently if at least one entry
of the coordinatewise thresholded point estimate $\hat {\boldsymbol \theta}({\lambda_\alpha})$  is different from zero has the desired level~$\alpha$.
Here $\lambda_\alpha=-\Phi^{-1}([1-\exp\{\log(1-\alpha)/N\}]/2)$ has a closed form expression, otherwise one can estimate it by Monte Carlo.
\cite{CandesANOVA:2011} call this test a max-test for an obvious reason.
 \item based on blockwise thresholding (\ref{eq:JS+}): likewise, for a sample ${\bf y}$, $\lambda_{\bf y}=\|{\bf y} \|_2^2$
 is the smallest and finite threshold that sets all the estimated parameters to zero.
 Under the null, that statistics $\Lambda_0 \sim \chi^2_N$. Therefore choosing the threshold $\lambda_\alpha=F_{\chi^2_N}^{-1}(1-\alpha)$,
 the test that rejects $H_0$ if $\hat {\boldsymbol \theta}({\lambda_\alpha})$ is different from the null vector leads to another test of level $\alpha$.
\end{itemize}
\cite{CandesANOVA:2011} studied the asymptotic relative power of both tests under either a dense or a sparse alternative hypothesis $H_1$.
According to their definition,  $H_1$ is dense or sparse whether the parameters ${\boldsymbol \theta}$ satisfy
either $\{ \|{\boldsymbol \theta} \|_2^2\geq B\}$ or $\{\|{\boldsymbol \theta} \|_\infty \geq A\}$ for some positive lower bounds $B$ and $A$, respectively.
They proved that the max-test has more power under the sparse alternative. %, under some assumptions of the design matrix.

Another goal of this paper is to derive thresholding-based tests for ANOVA considering either a coordinate or a block thresholding strategy, and see how they relate to and improve on existing tests.
%Since ANOVA is closely linked to linear models,
Our tests are based on thresholding estimators developed for linear models, and in particular
lasso \citep{Tibs:regr:1996}, grouped lasso \citep{Yuan:Lin:mode:2006} and smooth blockwise iterative thresholding \citep{SardySBITE2012}.
We show how the threshold parameter $\lambda_\alpha$ of these estimators can be determined for the thresholding test to have the desired level $\alpha$.
\citet{lassotestarchiv13} consider a sequential approach of  testing the significance of the successive lasso coefficients.

Section~\ref{eq:oneway} starts will the simple one-way ANOVA, derives a blockwise and coordinatewise tests, and investigate their relative power under a dense and a sparse alternative.
To take the best of both (dense and sparse) alternative worlds, Section~\ref{subsct:joint} derives a single $\alpha$-level test,
called O\&+ test, that is nearly as powerful as either the blockwise or the coordinatewise tests under both alternatives.
Section~\ref{sct:Tukey} is concerned with Tukey multiple comparisons test, and proves that it amounts to a coordinate thresholding test.
The latter has the advantage of having the exact desired level  not only in the balanced situation (like Tukey's), but also in the unbalanced one (where Tukey's is conservative),
albeit a Monte Carlo estimate of the threshold.
Section~\ref{sct:generalANOVA} presents the general framework for iterative thresholding-based tests for ANOVA, for which some coefficients may be thresholded coordinatewise, blockwise or both, depending
on the nature of the parameters (fixed effects, interactions, random effects) and on the nature of the alternative hypothesis considered (dense or sparse).
Section~\ref{sct:appli} applies the  new  test to  a real data set modeled by mixed effects.
Section~\ref{sct:YuanLin} proposes another selection of the threshold based on an extension of the universal threshold  to satisfy both
good estimation and model selection properties.
Section~\ref{sct:conclusion} draws some conclusions.

%%%%%%%%%%%%%%%%%%%%%%%%
%%%%%%%%%%%%%%%%%%%%%%%%
\section{One-way ANOVA: two tests}
\label{eq:oneway}

To fix notation, consider one-way ANOVA with $T$ treatments and $R$ replications
\begin{equation} \label{eq:oneway0}
Y_{tr}=\mu_t+\epsilon_{tr}, \quad t=1,\ldots,T, \ r=1,\ldots,R,
\end{equation}
where $\epsilon_{tr}\stackrel{\rm i.i.d.}\sim{\rm N}(0,\sigma^2)$
and the total number of observations is $N=TR$. In matrix notation, (\ref{eq:oneway0}) is ${\bf Y}=X {\boldsymbol \mu}+{\boldsymbol \epsilon}$,
where $X$ is an $N\times T$ matrix with
$$
{\bf Y}=\left (
\begin{array}{c}
 Y_{11}\\ \vdots\\ Y_{1R}\\  Y_{21}\\ \vdots\\ Y_{2R}\\ \vdots \\ Y_{T1}\\ \vdots \\ Y_{TR}
\end{array}
\right )
\quad
{\rm and}
\quad
X=\left (
\begin{array}{ccccc}
 1 & 0 & 0 & \ldots & 0 \\
 \vdots & \vdots & \vdots & \ldots & 0 \\
 1 & 0 & 0 & \ldots &   \vdots  \\
 0 & 1 & 0 & \ldots & 0 \\
 \vdots & \vdots  & \vdots &\ldots  & \vdots \\
 0      & 1        & 0 & \ldots & 0 \\
 \ldots & \ldots   & \ldots & \ldots & \ldots \\
 0 & 0 & 0 & \ldots & 1 \\
 \vdots & \vdots & \vdots & \ldots & \vdots \\
 0 & 0 & 0 & \ldots & 1 
\end{array}
\right ) .
$$
The matrix has orthogonal columns since $X\T X=R\ I_T$.
%Let also ${\boldsymbol \mu}=(\mu_1,\ldots,\mu_P)$ be the treatment effects.
For testing 
\begin{equation} \label{eq:H0mu}
H_0: \mu_1=\ldots=\mu_T (=\mu) \quad {\rm against} \quad H_1: \mbox{for at least one } t,\ \mu_t\neq \mu
\end{equation}
% \begin{equation} \label{eq:H0mu}
% \begin{array}{rl}
% H_0:& \mu_1=\ldots=\mu_T (=\mu) \\
% H_1:& \mbox{for at least one } $t$,\ \mu_t\neq \mu
% \end{array}
% \end{equation}
the most common approach is Fisher's test based on the pivot
$$
F_{{\bf Y}_0}=\frac{({\rm RSS}_{H_0}-{\rm RSS})/(T-1)}{{\rm RSS}/(N-T)} \sim {\cal F}_{T-1,N-T},
$$
where ${\cal F}_{d_1,d_2}$ is the Fisher distribution with degrees of freedom $d_1$ and $d_2$, ${\rm RSS}=\| {\bf Y}_0-X\hat{\boldsymbol \mu}^{\rm LS} \|_2^2$
with $\hat{\boldsymbol \mu}^{\rm LS}=(X\T X)^{-1}X\T {\bf Y}_0$, ${\rm RSS}_{H_0}=\| {\bf Y}_0-\bar Y_0 {\bf 1}\|_2^2$ and ${\bf Y}_0 \sim {\rm N}(\mu {\bf 1}, \sigma^2 I_N)$.
Next section shows that the same test can be derived based on block thresholding.

\subsection{Blockwise thresholding test}
\label{subsct:blocktest}

It is well known that
model~(\ref{eq:oneway0}) and test~(\ref{eq:H0mu}) are equivalent to testing
\begin{equation}\label{eq:H0alpha}
H_0: \theta_1=\ldots=\theta_T=0 \quad {\rm against} \quad H_1: \mbox{for at least one } t, \ \theta_t \neq 0
\end{equation}
% \begin{equation}\label{eq:H0alpha}
% \begin{array}{rl}
% H_0:& \theta_1=\ldots=\theta_T=0 \\
% H_1:& \mbox{for at least one } $t$, \ \theta_t \neq 0
% \end{array}
% \end{equation}
for the model
\begin{equation} \label{eq:oneway1}
{\bf y}=\mu {\bf 1}+X {\boldsymbol \theta}+{\boldsymbol \epsilon}.
\end{equation}
% \begin{equation} \label{eq:oneway1}
% y_{tr}=\mu+\theta_t+\epsilon_{tr}, \quad t=1,\ldots,T, \ r=1,\ldots,R.
% \end{equation}
As opposed to (\ref{eq:H0mu}), the formulation (\ref{eq:H0alpha}) of the null hypothesis is sparse in the sense that the parameters of the null hypothesis are all zero.
This motivates the following test based on thresholding.
For a given level $\alpha$, the idea is to derive a thresholded point estimate $\hat {\boldsymbol \theta}({\lambda_\alpha})$  and to control the threshold $\lambda_\alpha$
such that the point estimate is the null vector with the desired probability $1-\alpha$ under the null hypothesis.
The parameter $\mu$ is not to be tested, so we first calculate
\begin{equation} \label{eq:P_A}
{\bf y}_{A}={\bf y}-P_{A}{\bf y} \quad {\rm with} \quad P_A=A (A\T A)^{-1} A\T
\end{equation}
to remove the contribution of the $N \times 1$ matrix $A={\bf 1}$ corresponding to $\mu$. Here $P_{A}$ is the projection matrix in the range of $A$.
Then, given a positive threshold $\lambda$ and a smoothness parameter $s\geq 1$, the block threshold estimate \citep{SardySBITE2012} 
\begin{equation} \label{eq:SBITE4oneway}
\begin{array}{rcl}
  \hat {\boldsymbol \theta}(\lambda) &=& \left (1-\frac{\lambda}{ \|X\T {\bf y}_{A} \|_2} \right )_+^s \frac{1}{R} X\T {\bf y}_{A}
  \end{array}
\end{equation}
generalizes truncated James-Stein's thresholding (\ref{eq:JS+}) and 
% The parameter $s$ guarantees uniqueness of the point estimate in general ANOVA (see Section~\ref{sct:generalANOVA}) when $s>1$.
%obtained after one iteration of the algorithm.
has the property that $\hat {\boldsymbol \theta}(\lambda)={\bf 0}$ iff $\lambda \geq \|X\T {\bf y}_A \|_2$.
%with ${\bf y}_{A}={\bf y}-P_{A}{\bf y}$.
Note that $1/R$ in (\ref{eq:SBITE4oneway}) stems from the inverse of $X\T X$.
Observing that $\|X\T {\bf y}_A \|_2/\sigma$ is a pivot leads to the following theorem.

\bigskip
{\bf Theorem~1} (Block thresholding test): Consider model~(\ref{eq:oneway1}) for which we  test (\ref{eq:H0alpha})
%$H_0: {\boldsymbol \theta}={\bf 0}$  versus $H_1:\ {\boldsymbol \theta}\neq {\bf 0}$
at a prescribed level $\alpha$.
% with  $\epsilon_{tr}\stackrel{\rm i.i.d.}\sim{\rm N}(0,\sigma^2)$ and $\sigma$ known.
Let ${\bf Y}_0\sim {\rm N}(\mu {\bf 1}, \sigma^2 I_N)$ be the distribution of ${\bf Y}$ under $H_0$ and %let $\hat \sigma^2({\bf y}_0)={\rm RSS}({\bf y}_0)/(N-T)$ be the standard unbiased estimate of the variance.
%let  $\hat \sigma({\bf y}_0)$ be an estimate of $\sigma$ such that $\hat \sigma({\bf y}_0)/\sigma$ is a pivot.
let $\hat \sigma$ be a positive estimate of $\sigma$ such that  $\hat \sigma/\sigma$ is a pivot.
Then $\Lambda_{0,2}=\|X\T( {\bf Y}_0-P_A {\bf Y}_0) \|_2/\hat \sigma({\bf Y}_0)$ is a pivot with distribution $F^O_{\Lambda_{0,2}}$.
% Let $F_{\Lambda_{0,2}}$ be the distribution of the pivot, and let $\lambda_{\alpha,2}=F_{\Lambda_{0,2}}^{-1}(1-\alpha)$.
Defining the thresholded point estimate $\hat {\boldsymbol \theta}({\bf y}/\hat \sigma({\bf y}); {\lambda_{\alpha,2}})$
in (\ref{eq:SBITE4oneway}) for the observations ${\bf y}$ and setting the threshold $\lambda_{\alpha,2}$ such that $F^O_{\Lambda_{0,2}}(\lambda_{\alpha,2})=1-\alpha$, 
% for the data at hand ${\bf y}$ rescaled by $\hat \sigma({\bf y})$.
then the test
$$
\phi( {\bf y}) = \left \{
 \begin{array}{ll}
  1 & {\rm if} \ \hat {\boldsymbol \theta}({\bf y}/\hat \sigma({\bf y}); {\lambda_{\alpha,2}}) \neq {\bf 0} \\
  0 & {\rm otherwise} %{\rm if} \ \hat {\boldsymbol \alpha}_{\lambda_\alpha}({\bf y}/\hat \sigma({\bf y})) = {\bf 0} \\
 \end{array}
\right .
$$
has level $\alpha$.
Finally, letting $\hat \sigma^2({\bf y})$ %={\rm RSS}({\bf y}_0)/(N-T)$
be the standard unbiased estimate of variance,
the block thresholding test is equivalent to Fisher test with the relation $\lambda_{\alpha,2}^2=  R (T-1) F^{-1}_{{\cal F};T-1,N-T}(1-\alpha)$, where $F_{{\cal F}}$ is the cdf of the Fisher distribution.

{\bf Proof}: $\Lambda_{0,2}=(\| U_0 \|_2/\sigma)/(\hat \sigma({\bf Y}_0)/\sigma)$, where $U_0=X\T ({\bf Y}_0-P_A {\bf Y}_0) \sim {\rm N}({\bf 0}, \sigma^2(RI_T-R^2/N J_T))$
with $J_T$ the $T\times T$ matrix of ones.
%And $\hat \sigma^2({\bf y}_0)/\sigma^2 \sim \chi^2_{N-T}$.
So the distribution of ratio $\Lambda_{0,2}$ does not depend on $(\mu,\sigma)$. Moreover
$$
{\rm E}_{H_0} \phi( {\bf y})
= {\rm P}\left (\hat {\boldsymbol \theta}(\frac{{\bf Y}_0}{\hat \sigma({\bf Y}_0)}; {\lambda_{\alpha,2}}) \neq {\bf 0} \right )
={\rm P}\left (\frac{\| {\bf U}_0 \|_2}{ \hat \sigma({\bf Y}_0)} > \lambda_{\alpha,2} \right )
=1-F^O_{\Lambda_{0,2}}(\lambda_{\alpha,2})
=\alpha.
$$
The equivalence to Fisher's test using the standard unbiased estimate of variance for $\sigma^2$ is straightforward. $_\square$

\bigskip

By equivalence with Fisher's test, the distribution $F^O_{\Lambda_{0,2}}$ is known when using the standard unbiased estimate of variance,
so that the $(1-\alpha)$-quantile $\lambda_{\alpha,2}$ can be calculated from the quantile of Fisher's distribution.
In the more complex situations we will consider in the following, $F^O_{\Lambda_{0,2}}$ does not have an explicit expression.
This is for instance the case with this test if
another estimate of variance, say a robust one, possibly dependent of the numerator, is employed.
In that case $\lambda_{\alpha,2}$ can  be estimated by Monte-Carlo simulation.
%in a timely manner since simulating $\Lambda_0$ involves a Gaussian simulator and simple calculations.

%%%%%%%%%%%%%%%%%%%%%%%%%%%%%%%%%%%%%%
\subsection{Coordinatewise thresholding test}
\label{subsct:coordtest}

Instead of blocking the $T$ treatment parameters into one block of size $T$ and thresholding blockwise,
thresholding could be performed on $T$ blocks of size one, known as coordinatewise thresholding.
For blocks of size one, smooth blockwise iterative thresholding  \citep{SardySBITE2012} defines a point estimate 
as a solution to a set of nonlinear equations
\begin{equation} \label{eq:coordSBITE4oneway}
\left \{
\begin{array}{rcl}
   \hat \theta_t(\lambda) &=& \left (1-\frac{\lambda}{ |{\bf x}_t\T{\bf r}_t |} \right )_+^s \frac{1}{R} {\bf x}_t\T {\bf r}_t \quad {\rm with} \quad {\bf r}_t={\bf y}_A- \sum_{i \neq t} {\bf x}_i \hat \theta_i (\lambda)\\
   &&\quad t=1,\ldots,T 
  \end{array}
  \right .
\end{equation}
for a given positive threshold $\lambda$ and smoothness parameter $s\geq 1$.
Note that $1/R$ in (\ref{eq:coordSBITE4oneway}) stems from the inverse of ${\bf x}_t\T {\bf x}_t=R$ for all $t$.
For $s=1$, this defines the lasso estimate in a way that makes thresholding clearly visible: we recognize soft-thresholding (\ref{eq:soft})
applied to least squares estimates on the partial residuals.

% 
% Smooth blockwise iterative thresholding  \cite{SardySBITE2012} defines a point estimator capable thresholding either coordinatewise, blockwise or both.
% As opposed to lasso, group and adaptive lasso that are defined with an optimization problem, the point estimator
% %we derive another test  for (\ref{eq:H0alpha}) based on the following coordinate thresholded point estimate
%   is a fixed point to an iterative algorithm that thresholds least squares estimates, given the value of the other parameters.
Moreover the estimate $\hat {\boldsymbol \theta}(\lambda)=(\hat \theta_{1}(\lambda), \ldots, \hat \theta_{T}({\lambda}))$ satisfies the property that
$\hat \theta_{t}(\lambda)=0$ for all $t=1,\ldots,T$ iff $\lambda \geq \|X\T {\bf y}_A \|_\infty=\max_{t=1,\ldots,T} |{\bf x}_t\T {\bf y}_A|$.
%with ${\bf y}_A={\bf y}-P_A {\bf y}$.
This is a particular case of Lemma~1 proved in Section~\ref{sct:generalANOVA}.
Observing that $\|X\T {\bf y}_A \|_\infty/\sigma$ is a pivot leads to the following theorem, which proof is similar to that of Theorem~1.

%\bigskip
% 
% {\bf Property~2}: $\theta_{t,\lambda}({\bf y})=0$ for all $t=1,\ldots,T$ iff $\lambda \geq \|X\T {\bf y}-R\bar {\bf y} {\bf 1} \|_\infty$.
% 
% \bigskip

% This property is true for $s \geq 1$. It is a particular case of Property~4 proved below. Theorem~2 is also a particular case of Theorem~4 that addresses general ANOVA.

\bigskip
{\bf Theorem~2} (Coordinate thresholding test): Consider model~(\ref{eq:oneway1}) for which we  test (\ref{eq:H0alpha})
at a prescribed level $\alpha$.
Let ${\bf Y}_0\sim {\rm N}(\mu {\bf 1}, \sigma^2 I_N)$ be the distribution of ${\bf Y}$ under $H_0$ and 
let $\hat \sigma$ be a positive estimate of $\sigma$ such that  $\hat \sigma/\sigma$ is a pivot.
%let  $\hat \sigma({\bf y}_0)$ be an estimate of $\sigma$ such that $\hat \sigma({\bf y}_0)/\sigma$ is a pivot.
Then $\Lambda_{0,\infty}=\|X\T( {\bf Y}_0-P_A {\bf Y}_0) \|_\infty/\hat \sigma({\bf Y}_0)$ is a pivot with distribution $F^+_{\Lambda_{0,\infty}}$.
% 
% Consider model~(\ref{eq:oneway1}) for which we want to test $H_0: {\boldsymbol \theta}={\bf 0}$  versus $H_1:\ {\boldsymbol \theta}\neq {\bf 0}$ at a prescribed level $\alpha$.
% % with  $\epsilon_{tr}\stackrel{\rm i.i.d.}\sim{\rm N}(0,\sigma^2)$ and $\sigma$ known.
% Let ${\bf y}_0\sim {\rm N}(\mu {\bf 1}, \sigma^2 I_N)$ be the distribution under $H_0$ and let $\hat \sigma^2({\bf y}_0)={\rm RSS}({\bf y}_0)/(N-T)$
% be the standard unbiased estimate of the variance. Then $\Lambda_0=\|X\T {\bf y}_0-R\bar {\bf y}_0 {\bf 1} \|_\infty/\hat \sigma({\bf y}_0)$ is a pivot.
% Let $F_{\Lambda_{0,\infty}}$ be the distribution of the pivot, and let $\lambda_{\alpha,\infty}=F_{\Lambda_{0,\infty}}^{-1}(1-\alpha)$.
Defining the thresholded point estimate $\hat {\boldsymbol \theta}({\bf y}/\hat \sigma({\bf y}); {\lambda_{\alpha,\infty}})$
solution to (\ref{eq:coordSBITE4oneway}) for the observations ${\bf y}$ and setting the threshold $\lambda_{\alpha,\infty}$ such that
$F^+_{\Lambda_{0,\infty}}(\lambda_{\alpha,\infty})=1-\alpha$, 
then the test
$$
\phi( {\bf y}) = \left \{
 \begin{array}{ll}
  1 & \mbox{if} \ \hat {\theta}_{t}({\bf y}/\hat \sigma({\bf y});  \lambda_{\alpha,\infty}) \neq 0 \mbox{ for at least one } t \in \{1,\ldots,T\} \\
  0 & {\rm otherwise} %{\rm if} \ \hat {\boldsymbol \alpha}_{\lambda_\alpha}({\bf y}/\hat \sigma({\bf y})) = {\bf 0} \\
 \end{array}
\right .
$$
has level $\alpha$. 
% 
% \bigskip
% Note that $\Lambda_{0,\infty}$ is the max of what resembles $t$-statistics on each coefficient.

%%%%%%%%%%%%%%%%%%%%%%%%%%%%%%%%%%%%%%%%%%%%%%%%%%%%%%%%%%%%%%%%%
\subsection{Power analysis of both tests under two alternatives}
\label{subsct:bestboth}

We now compare the power of the two tests proposed in Sections~\ref{subsct:blocktest} and \ref{subsct:coordtest}. % under two alternatives.
%Assuming for simplicity that $\mu$ and $\sigma$ are known, 
Given a level $\alpha$, the thresholds
%We can do a comparative power analysis of both thresholding tests  by assuming that $\mu$ and $\sigma$ are known, say $\mu=0$ and $\sigma=1$,
%in which case we get a closed form expression for the power. %  without relying on Monte-Carlo simulation.
%For both thresholding tests, blockwise (based on the $2$-norm) and coordinatewise (based on the $\infty$-norm), to have the desired level~$\alpha$,
%the thresholds have the following closed form
\begin{equation}\label{eq:lambda2infty}
\lambda_{\alpha,2}=\sqrt{R} \sqrt{F_{\chi_T^2}^{-1}(1-\alpha)} \quad {\rm and} \quad \lambda_{\alpha,\infty}=- \sqrt{R} \Phi^{-1} \left (\frac{1-(1-\alpha)^{1/T}}{2} \right )
\end{equation}
respectively confer the blockwise (based on the $2$-norm) and coordinatewise (based on the $\infty$-norm) tests the desired level.
We assume here that $\mu$ and $\sigma$ are known for simplicity, but the conclusions below would hold if both $\mu$ and $\sigma$ were estimated.
%This means in particular that there is no need to use a projection matrix $P_A$, and so  ${\bf y}_A={\bf y}$ in (\ref{eq:P_A}).
 \cite{CandesANOVA:2011} prove the asymptotic result
that the test based on coordinate thresholding behaves differently than Fisher's test in terms of power depending whether the alternative hypothesis is dense or sparse.
% An alternative hypothesis is dense or sparse when the parameters are not null and whether they satisfy
% either $\{{\boldsymbol \theta}: \|{\boldsymbol \theta} \|_2^2\geq B\}$ or $\{{\boldsymbol \theta}: \|{\boldsymbol \theta} \|_\infty \geq A\}$, respectively.

%We investigate the exact power of both tests on a finite sample ANOVA with a realistic number of treatments and replications.
We consider two alternatives: a dense alternative of the form
$$
H_1^{D}: {\boldsymbol \theta}=\theta (\pm 1,\ldots,\pm 1),
$$
and a sparse alternative of the form
$$
H_1^{S}: {\boldsymbol \theta}=\theta (\pm 1,0,\ldots,0),
$$
where $\theta \in \real$.
The power of both tests as a function of $\theta$ under both alternatives is reported in Table~\ref{tab:power}.
Figure~\ref{fig:power1x2} plots power as a function of $\theta$ for $T=5$ treatments and $R=10$ replications.
This corroborates the asymptotic results of \cite{CandesANOVA:2011}:
%, in that the relative powers of both tests depends on the alternative hypothesis.
for a dense alternative, Fisher/block-test is more powerful, while for a sparse alternative
coordinate-test is more powerful.

\begin{table}[h!]  \caption{Power of blockwise and coordinatewise  thresholding tests at a level $\alpha$ under a dense and sparse alternative.
Notation: $\Delta_{\theta}\Phi(\lambda;R)=\Phi((\lambda-R \theta)/\sqrt{R})-\Phi((-\lambda-R \theta)/\sqrt{R})$.}
\label{tab:power}
  \begin{tabular}{l|cc} \hline
          & dense & sparse \\ \hline
block $\lambda=\lambda_{\alpha,2}$     & $1-F_{\chi^2_{T,RT \theta^2}}(\lambda^2/R)$ & $1-F_{\chi^2_{T,R\theta^2}}(\lambda^2/R)$ \\
coordinate $\lambda=\lambda_{\alpha,\infty}$ & $1-\{ \Delta_{\theta}\Phi(\lambda;R)\}^T$ &  $1-\Delta_{\theta}\Phi(\lambda;R) \{\Delta_{0}\Phi(\lambda;R)\}^{T-1}$\\
  \end{tabular}
\end{table}

%%%%%%%%%%%%%%%%%%%%%%%%%%%%%%%%%%%%
\section{One way ANOVA: the  O\&+ test}
\label{subsct:joint}

The relative power of both tests calls for a single test that would be of level~$\alpha$ while being as powerful as the best between the block- and coordinate-tests.
The following test approaches that goal by defining a point estimate based on joint block- and a coordinate-thresholding on the same parameters.

We now explain the notation $F^O$, $F^+$ and $F^{O\&+}$.
In dimension two, the $\ell_2$-ball  employed by Fisher block-test is a circle symbolized by 'O', the two canonical directions employed by coordinate thresholding are the horizontal
and vertical directions symbolized by '+', and so we call the joint test the O\&+ test.

To define the O\&+ test, consider the concatenated matrix $[X, X]$ and two sets of coefficients ${\boldsymbol \theta}_1$ and
${\boldsymbol \theta}_2$. Block ${\boldsymbol \theta}_1$ into one block and treat the second coordinatewise.
For reason we will explain, rescale the matrix $X$ into $X_1=X D_1$ and $X_2=X D_2$, where $D_1$ and $D_2$ are diagonal.
Finally consider the point estimate $\hat {\boldsymbol \theta}(\lambda)=(\hat {\boldsymbol \theta}_1({ \lambda}), \hat \theta_{2,1}({ \lambda}), \ldots, \hat \theta_{2,T}({ \lambda}))$ defined as a solution to 
\begin{equation} \label{eq:jointSBITE4oneway}
\left \{
\begin{array}{rcl}
   \hat {\boldsymbol \theta}_1(\lambda) &=& \left (1-\frac{\lambda}{ \|X_1\T{\bf r}_1 \|_2} \right )_+^s \frac{1}{R} D_1^{-2}X_1\T{\bf r}_1 \quad {\rm with} \quad {\bf r}_1={\bf y}_A - X_2 \hat {\boldsymbol \theta}_2(\lambda)\\
   \hat \theta_{2,t}(\lambda) &=& \left (1-\frac{\lambda}{ |{\bf x}_{2,t}\T {\bf r}_{2,t} |} \right )_+^s
   \frac{1}{Rd_{2,t}^2} {\bf x}_{2,t}\T {\bf r}_{2,t} \quad {\rm with} \\ 
   && \quad {\bf r}_{2,t}={\bf y}_A- X_1 {\boldsymbol \theta}_1-\sum_{i \neq t} {\bf x}_{2,i}  \hat \theta_{2,i}(\lambda) \quad t=1,\ldots,T 
  \end{array}
  \right .
\end{equation}
for a given positive threshold $\lambda$ and smoothness parameter $s \geq 1$, where ${\bf y}_A={\bf y}-P_A{\bf y}$ as before.
The solution to the system is unique if $s>1$ \citep{SardySBITE2012}, and has the property that 
% $\hat {\boldsymbol \theta}_{1,\lambda}({\bf y})=\hat {\boldsymbol \theta}_{2,\lambda}({\bf y})={\bf 0}$
$\hat {\boldsymbol \theta}({\lambda})={\bf 0}$
iff $\lambda \geq \max \{\|X_1\T {\bf y}_A  \|_2, \|X_2\T {\bf y}_A \|_\infty\}$ for $s\geq 1$. 
This is a particular case of Lemma~1 proved below, which leads to the following test.

%\bigskip

% {\bf Property~3}: $\hat {\boldsymbol \theta}_{1,\lambda}({\bf y})=\hat {\boldsymbol \theta}_{2,\lambda}({\bf y})={\bf 0}$ iff $\lambda \geq \max \{\|X_1\T {\bf y}-R D_1 \bar {\bf y} {\bf 1} \|_2, \|X_2\T {\bf y}-RD_2 \bar {\bf y} {\bf 1} \|_\infty\}$. \\
% {\bf Proof}: the implication is straightforward. For the converse,  ${\bf 0}$ is a solution. When $s>1$, the solution is unique
% \cite{SardySBITE2012}; when $s=1$, the solution is also \fbox{unique?}.

%\bigskip

% This property is true for $s \geq 1$. It is a particular case of Property~4 proved below. Theorem~3 is also a particular case of Theorem~4 that addresses general ANOVA.

\begin{figure}[!h]
  \begin{center}
  \includegraphics[height=12.5cm, width=9cm, angle=-90]{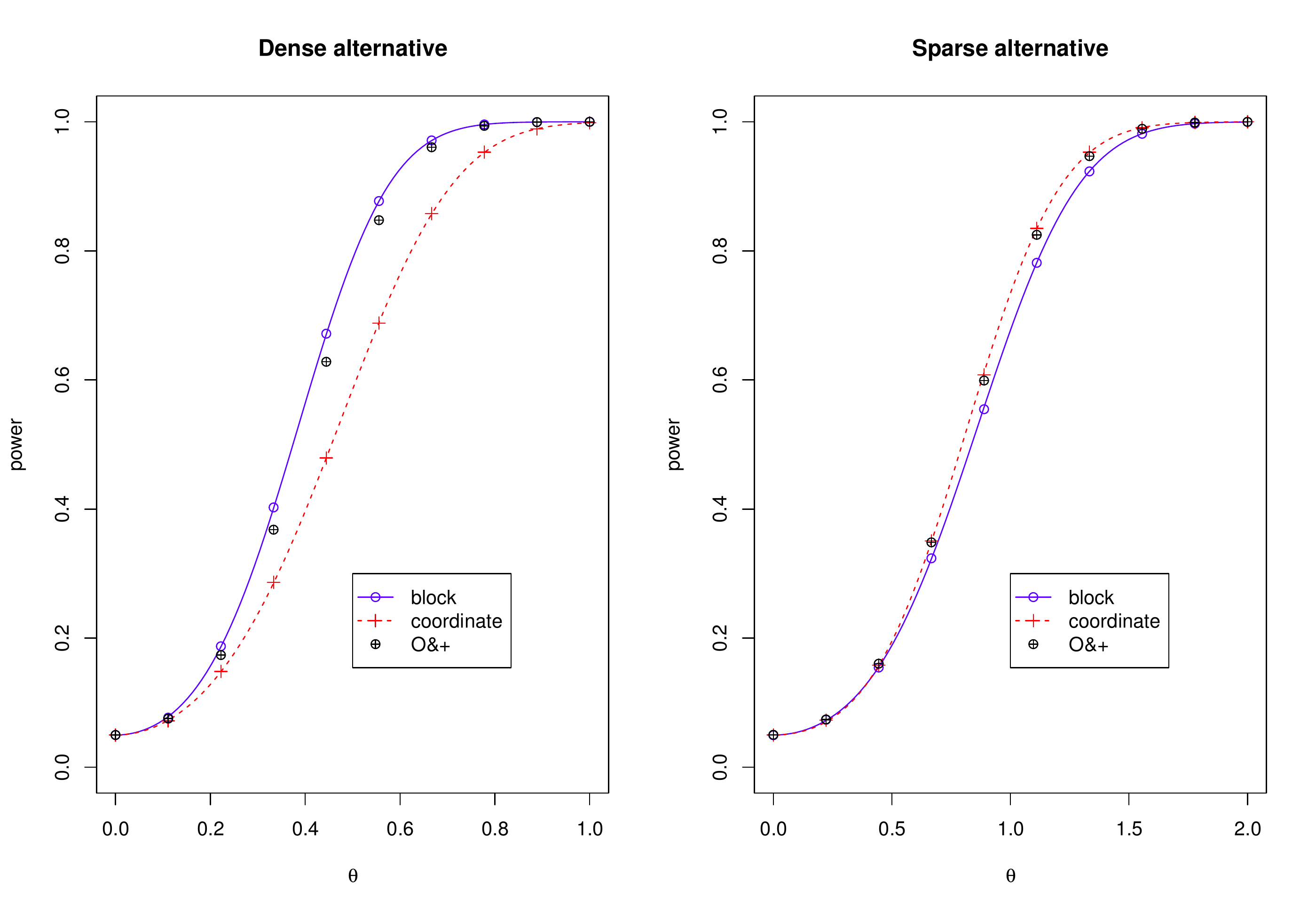}
  \caption{Power analysis of the three thresholding tests at a prescribed level $\alpha=.05$ for $T=5$ treatments and $R=10$ replications:
  block, coordinate and O\&+ tests.
  The lines are the theoretical powers of Table~\ref{tab:power}
    and the dots are empirical probabilities estimated by Monte-Carlo. Note that each test starts at power $\alpha=0.05$ for $\theta=0$, as expected.}
  \label{fig:power1x2}
  \end{center}
\end{figure}

\bigskip
{\bf Theorem~3} (O\&+ test): Consider model~(\ref{eq:oneway1}) for which we  test (\ref{eq:H0alpha})
at a prescribed level $\alpha$.
Let ${\bf Y}_0\sim {\rm N}(\mu {\bf 1}, \sigma^2 I_N)$ be the distribution of ${\bf Y}$ under $H_0$ and 
let $\hat \sigma$ be a positive estimate of $\sigma$ such that  $\hat \sigma/\sigma$ is a pivot.
%let  $\hat \sigma({\bf y}_0)$ be an estimate of $\sigma$ such that $\hat \sigma({\bf y}_0)/\sigma$ is a pivot.
% Then $\Lambda_{0,(2,\infty)}=\max \{ \|D_1(X\T {\bf y}_0-R\bar {\bf y}_0 {\bf 1}) \|_2/\hat \sigma({\bf y}_0), \|D_2(X\T {\bf y}_0-R\bar {\bf y}_0 {\bf 1}) \|_\infty/\hat \sigma({\bf y}_0)\}$
Then
\begin{equation} \label{eq:lambda2inf}
\Lambda_{0,(2,\infty)}=\max \{ \|X_1\T ({\bf Y}_0-P_A {\bf Y}_0) \|_2)/\hat \sigma({\bf Y}_0), \|X_2\T ({\bf Y}_0-P_A {\bf Y}_0) \|_\infty/\hat \sigma({\bf Y}_0)\} 
\end{equation}
is a pivot with distribution $F^\oplus_{\Lambda_{0,(2,\infty)}}$.
Defining the thresholded point estimate $\hat {\boldsymbol \theta}({\bf y}/\hat \sigma({\bf y}); {\lambda_{\alpha,(2,\infty)}})$
solution to (\ref{eq:jointSBITE4oneway}) for the observations ${\bf y}$ and setting the threshold $\lambda_{\alpha,(2,\infty)}$
such that $F^\oplus_{\Lambda_{0,(2,\infty)}}(\lambda_{\alpha,(2,\infty)})=1-\alpha$, 
then the test
% and let $\lambda_\alpha=F_{\Lambda_0}^{-1}(1-\alpha)$.
% Setting its threshold to $\lambda_\alpha$, calculate the point estimate 
% $\hat {\boldsymbol \theta}_{1,\lambda_\alpha}$
% and $\hat { \theta}_{2,t,\lambda_\alpha}$, $\ldots$, $\hat { \theta}_{2,T,\lambda_\alpha}$ solution to (\ref{eq:jointSBITE4oneway})
% for the data at hand ${\bf y}$ rescaled by $\hat \sigma({\bf y})$.
% Then the test
$$
\phi( {\bf y}) = \left \{
 \begin{array}{ll}
  1 & \mbox{if }  \hat {\boldsymbol \theta}({\bf y}/\hat \sigma({\bf y}); \lambda_{\alpha,(2,\infty)}) \neq {\bf 0} \\
  0 & {\rm otherwise} %{\rm if} \ \hat {\boldsymbol \alpha}_{\lambda_\alpha}({\bf y}/\hat \sigma({\bf y})) = {\bf 0} \\
 \end{array}
\right .
$$
has level $\alpha$. 

\bigskip

\begin{figure}[!h]
  \begin{center}
  \includegraphics[height=12.5cm, width=6cm, angle=-90]{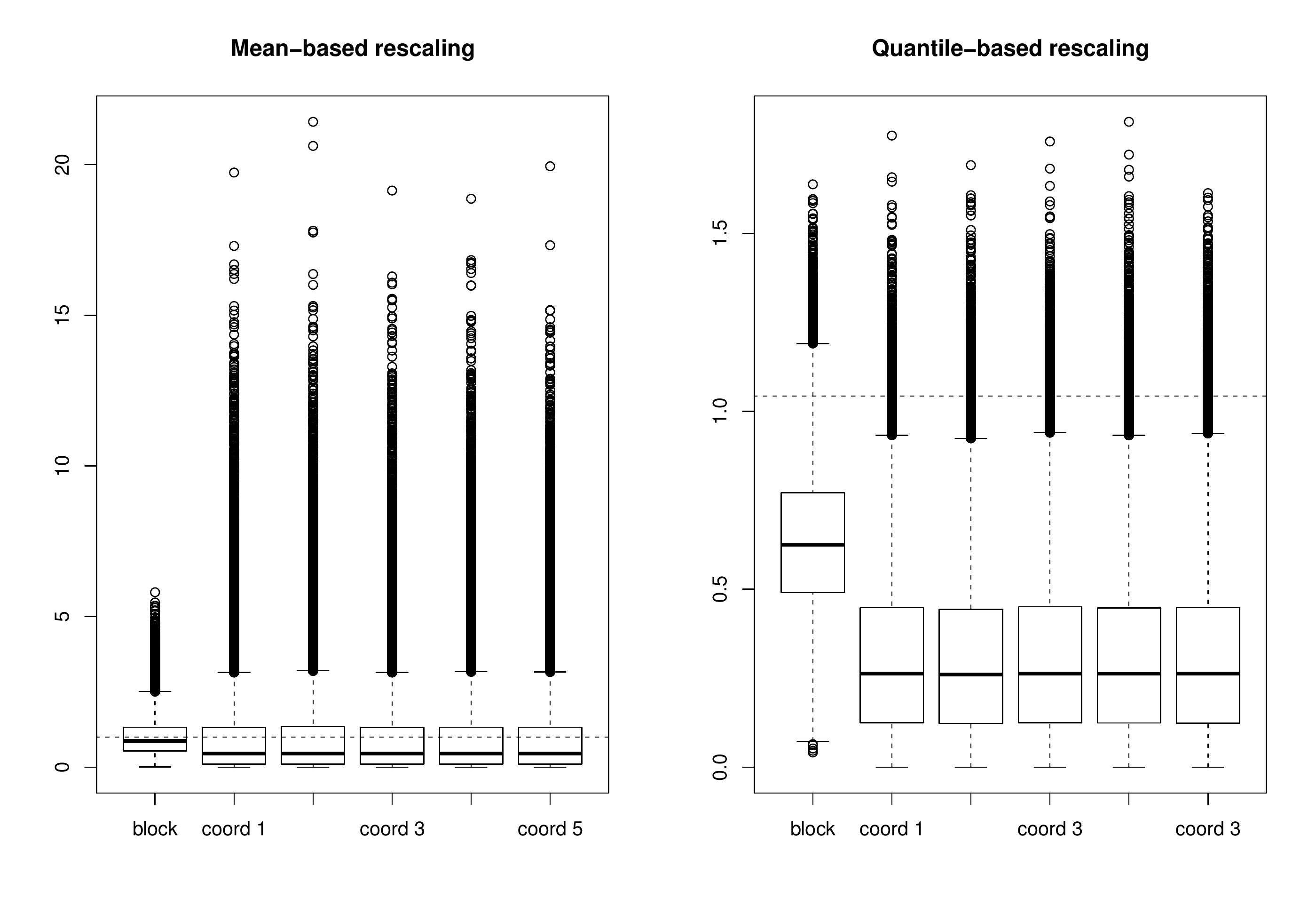}
    \caption{Illustration of mean-based rescaling (left) and quantile-based rescaling (right) for one-way ANOVA with $T=5$ treatments and $R=10$ replications.
    The series of 6 boxplots represents the empirical distribution of 
    the components of the pivot $\Lambda_{0,(2,\infty)}$ defined in (\ref{eq:lambda2inf}) under the null:
    %The quantile of the maximum of these distributions defines the threshold $\lambda_\alpha$ which provides a test at level $\alpha$.
   the first one corresponds to realizations of  $\|X_1\T {\bf Y}_0 \|_2$ and the other 5 correspond to realizations of $|{\bf x}_{2,t}\T {\bf Y}_0|$ for $t=1,\ldots,T$.
   We observe that the first boxplot on the right figure has the advantage of having upper extremes in the magnitude of the other five.}
   %This which is desired for the joint test to have good power under dense alternative.}
%   \caption{Illustration of mean-based rescaling (left) and quantile-based rescaling (right) for one-way ANOVA with $T=5$ treatments and $R=10$ replications. Each boxplot represents
%   the empirical distribution of the components of the pivot $\Lambda_{0,(2,\infty)}$ defined in (\ref{eq:lambda2inf}) under the null. The quantile of the maximum of these distributions defines the threshold $\lambda_\alpha$ which provides a test at level $\alpha$.
%   On each of the two plots, the left boxplot corresponds to realizations of  $\|X_1\T {\bf y}_0 \|_2$ and the other five correspond to realizations of $|{\bf x}_{2,t}\T {\bf y}_0|$ for $t=1,\ldots,T$.
%   On the right plot, the left boxplot has upper extremes in the magnitude of the other five, which is desired for the joint test to have good power under dense alternative.}
  \label{fig:boxplotH0}
  \end{center}
\end{figure}

Following on Section~\ref{subsct:bestboth} and Table~\ref{tab:power}
we consider the power of the joint-test under both dense and sparse alternatives as a function of $\theta$
(again assuming $\mu$ and $\sigma$ are known). %, so that ${\bf y}_A={\bf y}$ with $P_A=0$ in (\ref{eq:P_A})).
Figure~\ref{fig:power1x2} shows the power function of the block- and coordinate-tests (curve),
and estimate them as well  for values of $\theta$ on a grid with a Monte-Carlo simulation.
We also report the empirical power of the O\&+ test on the same grid, which has the remarkable property of performing under both alternatives almost as well as the best of the two individual tests.

To achieve with a single test a power nearly as good as the best of the two individual tests, rescaling the design matrix $X$ by $D_1$ and $D_2$ is a crucial step.
To allow the joint-test to have the same sensitivity whether the alternative is of the dense or sparse type, we perform the following quantile-based rescaling:
we let the matrix corresponding to the block coefficients ${\boldsymbol \theta}_1$ be $X_1=XD_1$ where $D_1$ is diagonal
with entries $1/\lambda_{\alpha,2}$; likewise we let  $X_2=XD_2$ where $D_2$ is diagonal with entries $1/\lambda_{\alpha,\infty}$
for the coefficients ${\boldsymbol \theta}_2$ thresholded coordinatewise.
The theoretical values of $\lambda_{\alpha,2}$ and $\lambda_{\alpha,\infty}$ are known (\ref{eq:lambda2infty}) in our simple setting,
but in more complex settings, we rely on a Monte-Carlo simulation to estimate them.
Figure~\ref{fig:boxplotH0} illustrates the advantage of quantile-based rescaling (right), as opposed to mean-based rescaling (left) proposed by \citet{Yuan:Lin:mode:2006} for group lasso.
% for the purpose of testing with the joint test.
The left plot shows that, under the null, the boxplots are centered around their means (horizontal dotted line); because of that rescaling,
the distribution of the block statistics $\|X_1\T {\bf Y}_0 \|_2$ (first boxplot from left) has its largest observations significantly lower than those of the coordinate statistics
$|{\bf x}_{2,1}\T {\bf Y}_0|, \ldots, |{\bf x}_{2,T}\T {\bf Y}_0|$ (second to sixth boxplots).
Consequently, with an alternative hypothesis of the dense type, the joint-test would have low power with that rescaling.
Instead, the right plot of Figure~\ref{fig:boxplotH0} shows how
quantile rescaling centers the distributions of the block and coordinate statistics
around their $(1-\alpha)$-quantile (horizontal dotted line), hence  providing the joint test with a homogeneous sensitivity under both dense and sparse alternatives.

%%%%%%%%%%%%%%%%%%%%%%%%%%%%%%%%%%%%%%%%
%%%%%%%%%%%%%%%%%%%%%%%%%%%%%%%%%%%%%%%%
\section{Tukey multiple comparisons test}
\label{sct:Tukey}

When the null hypothesis (\ref{eq:H0mu}) is rejected, \cite{tukey53} is interested in identifying which  null hypotheses
\begin{equation}\label{eq:H0MC}
H_0^{(t,t')}:\ \mu_t - \mu_{t'}=0, \quad t=1,\ldots,T, \ t'=t+1,\ldots,T
\end{equation}
 caused rejection of (\ref{eq:H0mu}).
%Consider the one-way layout of Section~\ref{eq:oneway}.
%With slightly more generality, Tukey allows the number of replication $R_t>0$ in treatment $t$ to not be all equal to $R$, 
%To test (\ref{eq:H0MC})
His test is dual to intervals
\begin{equation}\label{eq:TukeyInterval}
\bar y_t - \bar y_{t'} - \frac{{\cal T}_{T,N-T}(\alpha) \hat \sigma({\bf y})}{\sqrt{(R_t+R_{t'})/2}} \leq  \mu_t - \mu_{t'} \leq \bar y_t - \bar y_{t'} +  \frac{{\cal T}_{T,N-T}(\alpha) \hat \sigma({\bf y})}{\sqrt{(R_t+R_{t'})/2}}
\end{equation}
based on the Studentized range distribution ${\cal T}_{T,N-T}$, where $\hat \sigma^2({\bf y})$ is the unbiased estimate of variance.
Here $R_t>0$ is the number of replication in treatment $t$ (not necessarily all equal to $R$).
The test rejects $H_0^{(t,t')}$ if zero is not covered by its corresponding interval (\ref{eq:TukeyInterval}).
The level $\alpha$ of the test satisfies that  none of the $T(T-1)/2$ tests  are rejected with probability $1-\alpha$ under the null hypothesis (\ref{eq:H0mu}).
While the level is exact when $R_t=R$ for all $t$, the test is conservative \citep{Hayter84} for unbalanced designs, that is
when the number of replication $R_t$ differs between treatments.

% and  for a pair $(t,t')$ in all $T(T-1)/2$ pairs.
%The level of Tukey's test is $\alpha$ when $R_t=R$ for all $t$, and conservative otherwise \cite{Hayter84}.

Tukey's multiple comparisons test can also be derived based on thresholding and its level can be set to $\alpha$, even when the number of replication $R_t$ differs between treatments.
%An exact test with a level $\alpha$ can be derived with a thresholding test.
To see that, first note that the design matrix is orthogonal with $X^{\rm T}X={\rm diag}(R_1, \ldots, R_T)$.
Then let $E=(X^{\rm T} X)^{-1} X^{\rm T}$ and $\Delta$ be the $T(T-1)/2 \times T$ matrix
such that ${\boldsymbol \delta}=\Delta {\boldsymbol \mu}$ are the pairwise differences. 
% Left multiplying~(\ref{eq:oneway}) be $\Delta E$ implies
% $
% \Delta E {\bf y}= {\boldsymbol \delta} + \Delta E {\boldsymbol \varepsilon}.
% $
% The diagonal $D^2$ of the covariance matrix $\sigma^2 \Delta E E\T \Delta\T $   is not constant, unless $R_t=R$ for all $t=1,\ldots,T$.
% Standardization (i.e., homoscedasticity) can be achieved by left multiplying by $D^{-1}$ leading to the model
Left multiplying~(\ref{eq:oneway}) be $\Delta E$ implies
$
\tilde {\bf y}= X {\boldsymbol \delta} + \tilde {\boldsymbol \varepsilon}
$
with $X=I$, $\tilde {\bf y}=\Delta E {\bf y}$ and $\tilde {\boldsymbol \varepsilon} = \Delta E {\boldsymbol \varepsilon}$.

The coordinate thresholding estimate defined in (\ref{eq:coordSBITE4oneway}) can now be applied to that latter model. First rescaling must be performed:
the diagonal $D^2={\rm diag}(\sigma^2 \Delta E E\T \Delta\T )$ of the covariance matrix of $ \tilde {\boldsymbol \varepsilon}$ is not constant, unless $R_t=R$ for all $t=1,\ldots,T$.
So we standardize $X=I$ such that the marginals of $X\T \tilde {\bf y}$ are Gaussian with identical variance under $H_0$,
by multiplying $X$ by $D^{-1}$.
% leading to
%  \begin{equation} \label{eq:tildey}
%  \tilde {\bf y}= D^{-1} {\boldsymbol \delta} + \tilde {\boldsymbol \varepsilon} %\quad {\rm where} \quad
% %\left \{
% %\begin{array}{l}
% % \tilde {\bf y} = D^{-1} \Delta E {\bf y} \\
% % \tilde {\boldsymbol \varepsilon} =D^{-1} \Delta E {\boldsymbol \varepsilon}
% %\end{array}
% %\right . ,
% ,
% \end{equation}
% \begin{equation} \label{eq:tildey}
% \tilde {\bf y}= D^{-1} {\boldsymbol \delta} + \tilde {\boldsymbol \varepsilon} \quad {\rm where} \quad
% \left \{
% \begin{array}{l}
%  \tilde {\bf y} = D^{-1} \Delta E {\bf y} \\
%  \tilde {\boldsymbol \varepsilon} =D^{-1} \Delta E {\boldsymbol \varepsilon}
% \end{array}
% \right . ,
% \end{equation}
% where the marginal distributions of $D^{-1 }\tilde {\boldsymbol \varepsilon}$ are now identical.
Since the matrix $X=D^{-1}$  is diagonal, the coordinate thresholding estimate defined in (\ref{eq:coordSBITE4oneway})
has the closed form expression
% \begin{equation} \label{eq:deltahat}
%   \hat \delta_{i,\lambda} = \left (1-\frac{\lambda}{|\tilde y_i|/\hat \sigma} \right )_+ d_i \tilde y_i, \quad i=1,\ldots,T(T-1)/2,
% \end{equation}
\begin{equation} \label{eq:deltahat}
  \hat \delta_{(t,t')}(\lambda) = \left (1-\frac{\lambda}{|\tilde y_{(t,t')}/d_{(t,t')}|} \right )_+ d_{(t,t')} \tilde y_{(t,t')},
\end{equation}
for all pairs $(t,t')$. This thresholded point estimate leads to the following test.
% 
% 
% Theorem~4 below shows that the coordinate thresholding test is equivalent to Tukey's multiple comparison test when $R_t=R$ for all $t$, and has the exact desired level otherwise.
% %, as opposed to  Tukey multiple comparison test that is conservative in that case.
% The test is based on thresholding $\tilde {\bf y}$. 

\begin{figure}[!h]
  \begin{center}
  \includegraphics[width=2.8in,height=5in,angle=-90]{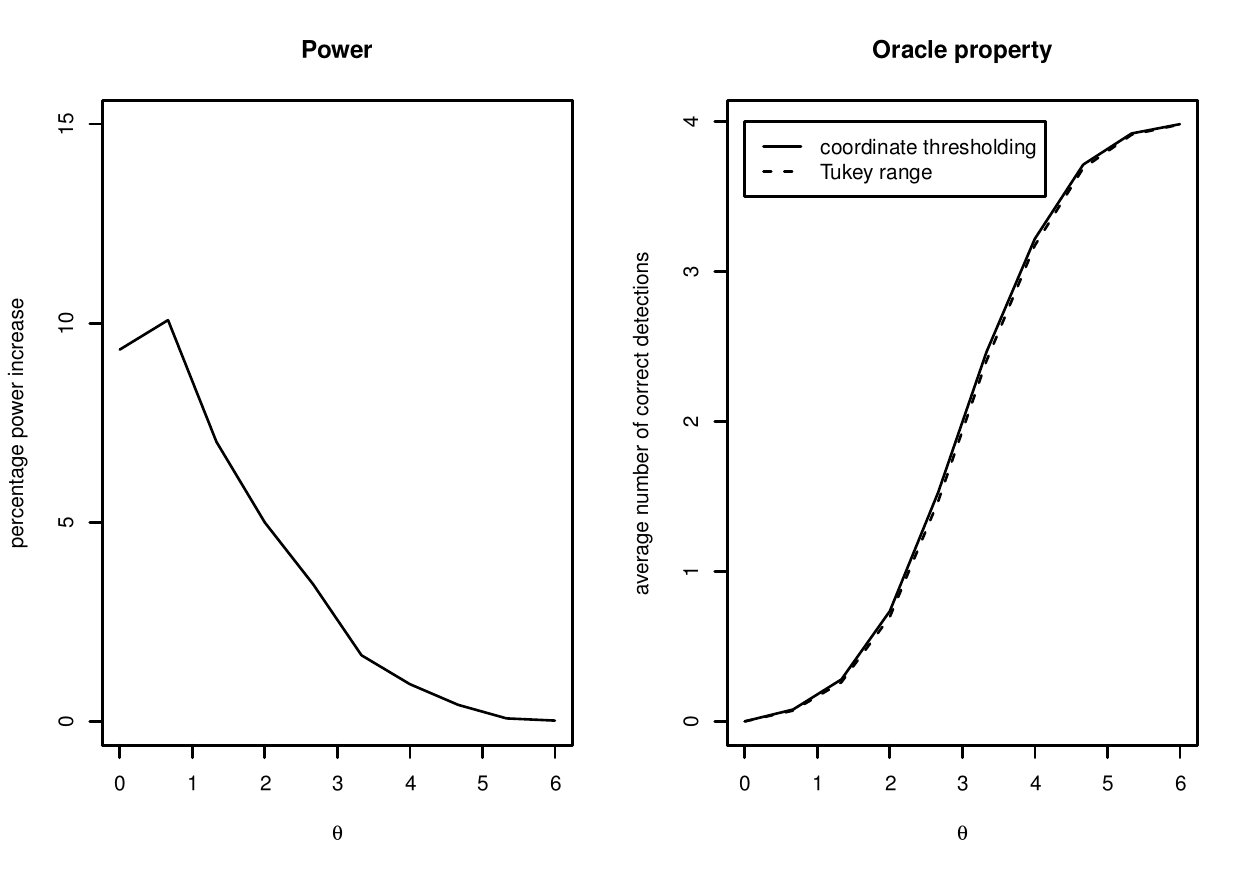}
  \caption{Results of Monte Carlo simulation for Tukey's multiple comparisons test for one-way ANOVA model (\ref{eq:oneway1})  with $T=5$ groups and sparse alternatives of the form $H_1:\ {\boldsymbol \theta}=(\theta,0,0,0,0)$
  with $\theta \in [0,6]$
  The number of replication is $(1,5,9,10,10)$ in each of the five groups.
  Left plot: percentage increase in power between the exact thresholding test (coordinate thresholding) and the conservative Tukey test.
  Right plot: average number of correct detections as a function of $\theta$, the maximum possible being $T-1=4$.}
  \label{fig:Tukeyrangerelative}
  \end{center}
\end{figure}

\bigskip
{\bf Theorem~4} (Coordinate thresholding test and Tukey multiple comparisons test).
Consider model~(\ref{eq:oneway0}) for which we  test all null hypotheses (\ref{eq:H0MC}) at a prescribed level $\alpha$.
Let $ {\bf Y}_0\sim {\rm N}(\mu {\bf 1}, \sigma^2 I_N)$ be the distribution of ${\bf Y}$ in (\ref{eq:oneway0})
%under $H_0$ for some $\mu \in \real$
and let $\hat \sigma^2$ %({\bf y}_0)={\rm RSS}({\bf y}_0)/(N-T)$
be the standard unbiased estimate of the variance. Then $\Lambda_{0,\infty}=\|D^{-1} \Delta E {\bf Y}_0 \|_\infty/\hat \sigma({\bf Y}_0)$ is a pivot
with distribution $F^+_{\Lambda_{0,\infty}}$.
Defining the thresholded point estimate $\hat {\boldsymbol \delta}({\bf y}/\hat \sigma({\bf y}); {\lambda_{\alpha, \infty}})$ in (\ref{eq:deltahat})
for the observations ${\bf y}$ and setting the threshold $\lambda_{\alpha,\infty}$ such that $F^+_{\Lambda_{0,\infty}}(\lambda_{\alpha,\infty})=1-\alpha$,
then the test
% 
% Let $F_{\Lambda_0}$ be the distribution of the pivot, and let $\lambda_\alpha=F_{\Lambda_0}^{-1}(1-\alpha)$.
% Setting its threshold to $\lambda_\alpha$, calculate the point estimate $\hat {\boldsymbol \delta}_{\lambda_\alpha}$ in (\ref{eq:deltahat}).
% for the data at hand ${\bf y}$ rescaled by $\hat \sigma({\bf y})$.
$$
\phi( {\bf y}) = \left \{
 \begin{array}{ll}
  1 & {\rm if}\ \hat \delta_{(t,t')}({\bf y}/\hat \sigma({\bf y}); \lambda_{\alpha,\infty}) \neq  0  \mbox{ for at least one } (t,t') \\
  0 & {\rm otherwise} %{\rm if} \ \hat {\boldsymbol \alpha}_{\lambda_\alpha}({\bf y}/\hat \sigma({\bf y})) = {\bf 0} \\
 \end{array}
\right .
$$
has level $\alpha$.
Moreover if $R_t=R$ for all treatments, this test is equivalent to Tukey's range test with the relation ${\cal T}_{T,N-T}(\alpha)=\lambda_{\alpha,\infty} \sqrt{2}$, where $N=R T$.

{\bf Proof}: the proof that this test has level $\alpha$ is straightforward. For the equivalence to Tukey multiple comparisons test when $R_t=R$ for all treatments~$t$,
note that on the one hand Tukey's test rejects when at least one interval (\ref{eq:TukeyInterval}) does not cover zero, or equivalently when
$|\bar y_t - \bar y_{t'}| - {\cal T}_{T,N-T}(\alpha) \hat \sigma({\bf y})/\sqrt{R}\leq 0$ for at least one pair $(t,t')$. 
On the other hand, the thresholding test rejects when $|\tilde y_{(t,t')}/d_{(t,t')}/\hat \sigma({\bf y})|\leq  \lambda_{\alpha,\infty}$, where $\tilde y_{(t,t')}=(\Delta E {\bf y})_{(t,t')}=\bar y_t - \bar y_{t'}$,
and  all $d_{(t,t')}^2=2/R$ when $R=R_t$.
So the thresholding test rejects when $|\bar y_t - \bar y_{t'}| \leq \lambda_{\alpha,\infty} \hat \sigma({\bf y}) \sqrt{2/R}$.
So the two tests are equivalent and ${\cal T}_{T,N-T}(\alpha)=\lambda_{\alpha,\infty} \sqrt{2}$. $_\square$

%\fbox{A revoir/multiple comparison/divided by sigma.../reformulate in $(t,t')$?/Proof?}
% \begin{figure}[!ht]
%   \begin{center}
%   \includegraphics[height=12cm, width=10cm, angle=-90]{Figures/Tukeyrange}
%   \caption{Monte Carlo Tukey range.}
%   \label{fig:Tukeyrange}
%   \end{center}
% \end{figure}

We illustrate the gain in power and oracle property of the thresholding test in comparison to Tukey's multiple comparisons test on the same Monte Carlo simulation as in Section~\ref{subsct:joint},
except that the number of replication in each of the $T=5$ treatments is $(1,5,9,10,10)$ instead of $R_t=R=10$ for all treatments.
The alternative hypothesis of the form $H_1:\ {\boldsymbol \theta}=(\theta,0,0,0,0)$
is indexed by the parameter $\theta$ in the range $[0,6]$.
In that setting, Tukey's test is conservative as we observe on the left plot of Figure~\ref{fig:Tukeyrangerelative} where the percentage increase of power is plotted as a function of $\theta$.
The right plot informs on the number of correct detections, with a target value of $T-1$ non-zero entries guessed correctly.
We see on the graph that the thresholding test has slighlty more correct detections  on average  than Tukey's conservative test.

If it is not clear to the statistician whether the alternative hypothesis is sparse or dense, then Tukey multiple comparisons test could be turned into
an O\&+ test as well to have more power under both alternatives.

Finally we considered pairwise contrasts, but the thresholding test could easily be implemented to test other contrasts.

%%%%%%%%%%%%%%%%%%%%%%%%%%%%%%%%%%%%%%%%%%%%%
\section{Thresholding test for general ANOVA models}
\label{sct:generalANOVA}

We now consider general ANOVA models which can be written in the form
\begin{equation} \label{eq:genANOVA}
{\bf y}=A {\bf b} + X {\boldsymbol \theta} + \sigma {\bf z} \quad {\rm with} \quad \left \{ \begin{array}{l} X=[X_1 \ldots X_Q], \\ 
{\boldsymbol \theta} = ({\boldsymbol \theta}_1, \ldots, {\boldsymbol \theta}_Q),  \\ {\bf z} \sim {\rm N}({\bf 0}, I_N). \end{array} \right .
\end{equation}
The matrix $A$ corresponds to the nuisance parameters~${\bf b}$. In the one-way ANOVA considered in the previous section, $A={\bf 1}$ and $b$ is the intercept coefficients, but $A$ can include 
a large number of parameters we do not want to test.
The $N \times P$ matrix $X$ corresponds to the parameters of interest, and is the horizontal concatenation of matrices $X_1, \ldots, X_Q$ corresponding to effects ${\boldsymbol \theta}_1, \ldots, {\boldsymbol \theta}_Q$, each
of respective size $P_q$ such that $\sum_{q=1}^Q P_q=P$.
For instance in a one-way ANOVA plus random effects, ${\boldsymbol \theta}_1$ are the main and ${\boldsymbol \theta}_2$ random effects, as in the application of Section~\ref{sct:appli}.
Another example of concatenated matrices is the joint test of Section~\ref{subsct:joint} with coefficients $({\boldsymbol \theta}_1, \theta_{21} ,\ldots, \theta_{2T})$
with corresponding matrices $X=[X_1 {\bf x}_{21} \ldots {\bf x}_{2T}]$.
%with one block of size $T$ for ${\boldsymbol \theta}_1$ and $T$ blocks of size one for $ \theta_{21} ,\ldots, \theta_{2T}$.
We assume that matrices $A$ and $X_1, \ldots, X_Q$ are all full rank; $X$ needs not be full rank.
Moreover we assume the space of the nuisance parameters are not too large, in the sense that all $X_q$ must have some components outside the range of $A$.

Our goal is to test the null hypothesis (\ref{eq:H0alpha}), that is test
\begin{equation} \label{eq:ANOVAtheta}
H_0:\ {\boldsymbol \theta}_1={\bf 0}, \ \ldots, {\boldsymbol \theta}_Q={\bf 0}
\end{equation}
at a desired level $\alpha$. To derive a threshold-based test, we must define a point estimate that thresholds and is uniquely defined.
\citet{SardySBITE2012} guarantees uniqueness of a thresholding estimator with a linear and invertible reparametrization of (\ref{eq:genANOVA}) into
\begin{equation} \label{eq:genANOVAgamma}
{\bf y}=A {\bf b} + \tilde X {\boldsymbol \gamma} + \sigma {\bf z} \quad {\rm with} \quad \left \{ \begin{array}{l} \tilde X=[\tilde X_1 \ldots \tilde X_Q], \\
{\boldsymbol \gamma} = ({\boldsymbol \gamma}_1, \ldots, {\boldsymbol \gamma}_Q),  \\ {\bf z} \sim {\rm N}({\bf 0}, I_N), \end{array} \right . 
\end{equation}
where each $\tilde X_q$ must satisfy the orthogonality condition: $\tilde X_q\T \tilde X_q=d_q^2 I_{P_q}$ with $d_q>0$.
Group lasso is also defined with this condition \citep{Yuan:Lin:mode:2006}, but not necessarily uniquely.
Orthogonalization can be achieved with a QR or SVD decomposition, for instance.
Testing  (\ref{eq:ANOVAtheta}) is equivalent to testing
\begin{equation} \label{eq:ANOVAgamma}
H_0:\ {\boldsymbol \gamma}_1={\bf 0}, \ \ldots, {\boldsymbol \gamma}_Q={\bf 0}.
\end{equation}
%\subsection{Deriving a test of level $\alpha$}
% 
% Orthogonalization and rescaling of all $X_q$ means a linear and invertible change of variables: the model is now parametrized in ${\boldsymbol  \gamma}_q$.
% Consequently the null hypothesis  $H_0: \ {\boldsymbol  \theta}_q={\bf 0}$ is equivalent to the null hypothesis $H_0:\ {\boldsymbol  \gamma}_q={\bf 0}$.
For a given threshold $\lambda > 0$ and a smoothness parameter $s\geq1$, we introduce the thresholded point estimate $\hat {\boldsymbol \gamma}$ defined as a solution (not necessarily unique unless $s>1$) to the following nonlinear system
\begin{equation}\label{eq:gammahat}
\left \{
\begin{array}{lll}
  \hat {\boldsymbol \gamma}_q(\lambda )&=&\left (1-\frac{\lambda}{ \| \tilde X_q^{\rm T} {\bf r}_q \|_2} \right )^s_+ \tilde X_q^{\rm T} {\bf r}_q/d_q \\ 
  && {\rm with}  \quad {\bf r}_q = {\bf y}_A - \sum_{q' \neq q} \tilde X_{q'} \hat {\boldsymbol \gamma}_{q'}(\lambda), \quad q\in {\cal Q}_{\rm block}   \\
  \hat \gamma_{q,t}(\lambda)&=&\left (1-\frac{\lambda}{ | \tilde {\bf x}_{q,t}^{\rm T} {\bf r}_{q,t} |} \right )^s_+ \tilde {\bf x}_{q,t}^{\rm T} {\bf r}_{q,t}/d_q \\
  && {\rm with}  \quad  {\bf r}_{q,t} = {\bf y}_A - \tilde {\bf X}_{-(q,t)} \hat  {\boldsymbol \gamma}_{-{q,t}}(\lambda), \quad q\in {\cal Q}_{\rm coord}, \ t=1,\ldots,P_q 
  \end{array}
\right . 
\end{equation}
where ${\bf y}_A$ is defined in (\ref{eq:P_A}) and ${\cal Q}_{\rm block}$ are the indexes of blocked variables (resp., ${\cal Q}_{\rm coord}$ for variables thresholded coordinatewise), and ${\bf X}_{-(q,t)}$ is the matrix $X$ without column $t$ of block $q$
\citep{SardySBITE2012}.
This point estimate has the following important property. % provided all blocks $X_q$ are full rank.

{\bf Lemma~1}: Considering  system (\ref{eq:gammahat}) for given $s\geq 1$ and $ \lambda>0$, then
\begin{equation} \label{eq:equiv1}
 \hat {\boldsymbol \gamma}_{q}(\lambda)={\bf 0} \ \mbox{for all} \ q \in {\cal Q}_{\rm block} \cup {\cal Q}_{\rm coord} 
\end{equation}
if and only if
\begin{equation} \label{eq:lambdayAX}
\lambda \geq \max \{ \max_{q\in {\cal Q}_{\rm block }} \|\tilde X_q^{\rm T} {\bf y}_A \|_2, \max_{q\in {\cal Q}_{\rm coord }} \|\tilde X_q^{\rm T} {\bf y}_A \|_\infty \}.
\end{equation}
In this case, $\hat {\boldsymbol \gamma}(\lambda)$ is uniquely defined.

{\bf Proof}: the implication is straightforward. For the converse, the choice of $\lambda$ in (\ref{eq:lambdayAX}) implies ${\bf 0}$ is a solution.
The proof is complete if the zero solution  is unique.
When $s>1$,  \citet{SardySBITE2012} proved uniqueness of the solution to (\ref{eq:gammahat}).
When $s=1$,  (\ref{eq:gammahat}) are the first order optimality conditions to the hybrid lasso-grouped lasso defined as solution to
$$
\min_{{\boldsymbol \gamma}} \frac{1}{2} \| {\bf y}_A - \tilde X {\boldsymbol \gamma} \|_2^2 + \lambda \| {\boldsymbol \gamma}\|_{\cal Q},
$$
where $\| {\boldsymbol \gamma}\|_{\cal Q}=\sum_{q\in {\cal Q}_{\rm block}} \|{\boldsymbol \gamma}_q\|_2 + \sum_{q\in {\cal Q}_{\rm coord}} \|{\boldsymbol \gamma}_q\|_1$ is a hybrid-norm.
This cost function $C({\boldsymbol \gamma})$ of the above minimization problem is convex in ${\boldsymbol \gamma}$, and the solution  may not be unique in that case.
In fact if two solutions ${\boldsymbol \gamma}_1$ and ${\boldsymbol \gamma}_2$  exist then their convex combinations 
${\boldsymbol \gamma}_\delta=\delta {\boldsymbol \gamma}_1+(1-\delta){\boldsymbol \gamma}_2$ are also solutions, that is
$C({\boldsymbol \gamma}_1)=C({\boldsymbol \gamma}_2)=C({\boldsymbol \gamma}_\delta)=C^*$ for all $\delta \in [0,1]$.
But suppose $\tilde X {\boldsymbol \gamma}_1\neq \tilde X {\boldsymbol \gamma}_2$, then the strict convexity of $\| \cdot \|_2^2$
and convexity of the ${\cal Q}$-norm imply that
\begin{eqnarray*}
 C({\boldsymbol \gamma}_\delta) &=& \frac{1}{2} \|\delta ({\bf y}_A - \tilde X {\boldsymbol \gamma}_1 ) + (1-\delta) ({\bf y}_A - \tilde X {\boldsymbol \gamma}_2 ) \|_2^2
 + \lambda \| \delta {\boldsymbol \gamma}_1 + (1-\delta) (\boldsymbol \gamma)_2\|_{\cal Q}\\
 &<&  \frac{1}{2} \{\delta \| {\bf y}_A - \tilde X {\boldsymbol \gamma}_1 \|_2^2  + (1-\delta) \| {\bf y}_A - \tilde X {\boldsymbol \gamma}_2 \|_2^2\}
 + \lambda (\delta \| {\boldsymbol \gamma}_1\|_{\cal Q} + (1-\delta)  \| {\boldsymbol \gamma}_1\|_{\cal Q})\\
 &=&C^*.
\end{eqnarray*}
This contradiction implies that any two solutions must satisfy $\tilde X {\boldsymbol \gamma}_1=\tilde X {\boldsymbol \gamma}_2$,
%but the strict convexity of $\| \cdot \|_2^2$ implies that any two solutions can only differ by elements in the kernel of $\tilde X$.
and have the same least squares cost. So their penalty term must be equal: $\lambda\| {\boldsymbol \gamma}_1\|_{\cal Q}=\lambda\| {\boldsymbol \gamma}_2\|_{\cal Q}$.
Since ${\boldsymbol \gamma}_1={\boldsymbol 0}$ is a solution, its ${\cal Q}$-norm is zero.
% So any other solution ${\boldsymbol \gamma}_2$ must satisfy $\| {\boldsymbol \gamma}_2\|_{\cal Q}=0$. Since $\| \cdot \|_{\cal Q}$
% is a norm, it implies that ${\boldsymbol \gamma}_2={\bf 0}$.
Then necessarily ${\boldsymbol \gamma}_2={\bf 0}$. $_\square$

\bigskip

The following theorem proposes a test of level $\alpha$ and shows it is a thresholding test since it amounts to testing whether a thresholded point estimate is null or not. Its proof is based on Lemma~1. 

\bigskip
{\bf Theorem~4} (Thresholding test for ANOVA):  Consider model~(\ref{eq:genANOVA}) for which we  test (\ref{eq:ANOVAtheta}) at a prescribed level $\alpha$.
% with  $\epsilon_{tr}\stackrel{\rm i.i.d.}\sim{\rm N}(0,\sigma^2)$ and $\sigma$ known.
Assume $A$ is full column rank and let $P_A$ be the projection in the range of $A$.
Let ${\bf Y}_0\sim {\rm N}(A{\bf b}, \sigma^2 I_N)$ be the distribution of ${\bf Y}$ under $H_0$ and let $\hat \sigma^2$ %={\rm RSS}({\bf y}_0)/(N-T)$
be the standard unbiased estimate of variance.
Then
\begin{equation}  \label{eq:pivotANOVA}
\Lambda_0= \max \{ \max_{q\in {\cal Q}_{\rm block }} \|\tilde X_q^{\rm T} ({\bf Y}_0-P_A {\bf Y}_0) \|_2, \max_{q\in {\cal Q}_{\rm coord }} \|\tilde X_q^{\rm T}  ({\bf Y}_0-P_A {\bf Y}_0) \|_\infty \}/\hat \sigma({\bf Y}_0)
\end{equation}
is a pivot with distribution $F_{\Lambda_0}$. Letting $\lambda_\alpha=F_{\Lambda_0}^{-1}(1-\alpha)$,
then the test
$$
\phi( {\bf y}) = \left \{
 \begin{array}{lll}
  1 & \mbox{if} \ \lambda_\alpha < \max \{ & \max_{q\in {\cal Q}_{\rm block }} \|\tilde X_q^{\rm T} ({\bf y}-P_A {\bf y}) \|_2 /\hat \sigma({\bf y}),\\
  &&\max_{q\in {\cal Q}_{\rm coord }} \|\tilde X_q^{\rm T}  ({\bf y}-P_A {\bf y}) \|_\infty /\hat \sigma({\bf y}) \} \\
  0 & {\rm otherwise} %{\rm if} \ \hat {\boldsymbol \alpha}_{\lambda_\alpha}({\bf y}/\hat \sigma({\bf y})) = {\bf 0} \\
 \end{array}
\right .
$$
has level $\alpha$.
Moreover defining the thresholded point estimate $\hat {\boldsymbol \gamma}({\bf y}/\hat \sigma({\bf y}); {\lambda_\alpha})$ as a solution to (\ref{eq:gammahat})
for the observation ${\bf y}$, then the test
$$
\tilde \phi( {\bf y}) = \left \{
 \begin{array}{ll}
  1 & \mbox{if} \ \hat {\boldsymbol \gamma}({\bf y}/\hat \sigma({\bf y});\lambda_\alpha) \neq {\bf 0}  \\
  0 & {\rm otherwise} %{\rm if} \ \hat {\boldsymbol \alpha}_{\lambda_\alpha}({\bf y}/\hat \sigma({\bf y})) = {\bf 0} \\
 \end{array}
\right . = \phi({\bf y}).
$$

%Not reject iff $\lambda \geq \max$ iff zero is a solution iff the set of solution is the zero singleton.

%Reject iff $\lambda < \max$ iff the set of solution is not the zero singleton iff there exists a solution not equal to zero.

\bigskip 

Rescaling the block matrices $X_q$ after orthonormalizing them is a crucial step as we illustrated in Section~\ref{subsct:joint}.
Quantile rescaling allows the $Q$ blocks to contribute equally to the distribution of the pivot $\Lambda_0$ in (\ref{eq:pivotANOVA}), regardless of their sizes.
% Consequently the test has a good power under various alternative hypothesis.
% 
% Also considering each block $q=1,\ldots,Q$, we orthonormalize it into a matrix $W_q$ (for instance using QR, SVD)
% and rescale it to $\tilde X_q$ such that $\tilde X_q\T \tilde X_q=d_q^2 I_{P_q}$ with $d_q>0$.
% The choice of $d_q$ is based on quantile rescaling.
% 
Quantile rescaling is defined as follows. Given a block $q$ and the orthonormalization $W_q$ of $X_q$ (with QR or SVD),
quantile rescaling applies the same factor to $W_q$ and leads to the rescaled matrix:
\begin{itemize}
 \item $\tilde X_q=W_q d_q$, where $1/d_q=\lambda_{\alpha,2}^{(q)}$ is the $(1-\alpha)$-quantile of the distribution
 of $\|W_q\T {\bf y}_A \|_2$ under the null,
 if the corresponding coefficients ${\boldsymbol \theta}$ are thresholded blockwise;
 \item $\tilde X_q=W_q d_q$, where $1/d_q=\lambda_{\alpha,\infty}^{(q)}$  is the $(1-\alpha)$-quantile of the distribution
 of $\|W_q \T {\bf y}_A \|_\infty$ under the null, if the corresponding coefficients ${\boldsymbol \theta}$ are thresholded coordinatewise.
\end{itemize}

In the case $P>N$, we propose the following estimate of the standard deviation $\sigma$. Let $X=U D V\T$ be the singular value decomposition of $X$, the design matrix of rank $R\leq N$.
Then $\hat {\boldsymbol  \gamma}^{\rm LS}=D \hat {\boldsymbol  \beta}^{\rm LS}\sim {\rm N}({\boldsymbol  \gamma}, \sigma^2 I_R)$, where  $\hat {\boldsymbol  \beta}^{\rm LS}$ %=D^{-2}V \T X\T {\bf y}$
is the least squares estimate with the transformed matrix $XV$, the matrix of principal component regression.
In eigen directions of small singular values $d_r$, the true coefficients $\gamma_r$ should essentially be zero.
So we propose to estimate the standard deviation with $\hat \sigma={\rm MAD}(|\hat {  \gamma}_{p_0}^{\rm LS}|, \ldots, |\hat {  \gamma}_{R}^{\rm LS}|)$ for $p_0$ large enough, say $p_0=\lfloor R/2 \rfloor$.
If $P$ is prohibitively large to prevent an SVD, then its columns can be sampled to create sample matrices of a reasonable size, and repeated estimations of $\sigma$ can then be aggregated into one.
Note that this resampling procedure should be reproduced under the null to determine the appropriate threshold.

%%%%%%%%%%%%%%%%%%%%%
%%%%%%%%%%%%%%%%%%%%%
\section{Application}
\label{sct:appli}

%\subsection{???}

%\subsection{Mixed effects model}

We illustrate the thresholding test on a real data set modeled with mixed-effects \citep{MixedEffectPinheiro}.
The effort $y_{ij}$ required (on the Borg scale) to arise from a stool is measured for $J=9$ different
subjects each using $I=4$ different types of stools.
A linear mixed-effects model can be written as
$$
y_{ij}=\theta_0 + X {\boldsymbol \theta}_1 + Z {\boldsymbol \theta}_2 + \epsilon_{ij},
$$
where $X$ model the fixed effects for Types (4 columns) and $Z$ models the random effect for Subjects (9 columns). The noise is assumed i.i.d.~${\rm N}(0,\sigma^2)$ and 
${\boldsymbol \theta}_2$ is believed to be independent realizations from ${\rm N}(0, \sigma_2)$. The goal is to test 
$$
H_0:\ \theta_{1,1}=\theta_{1,2}=\theta_{1,3}=\theta_{1,4}=0 \quad {\rm and} \quad \sigma_2=0,
$$
or equivalently
$$
H_0:\ \theta_{1,1}=\theta_{1,2}=\theta_{1,3}=\theta_{1,4}=0 \quad {\rm and} \quad {\boldsymbol \theta}_2={\bf 0}.
$$
% Testing $ \sigma_2=0$ amounts to blockwise thresholding the random effects ${\boldsymbol \theta}_2$ and testing whether the block is null or not.
% So we choose to threshold  the fixed effects ${\boldsymbol \theta}_1$ coordinatewise and the random effects ${\boldsymbol \theta}_2$ blockwise.
% For the fixed effects ${\boldsymbol \theta}_1$, we choose coordinatewise thresholding to determine which types of stool seem significant.
So we employ the thresholded test coordinatewise for the fixed effects and blockwise for the random effect.  Table~\ref{tab:mixed} reports the result  at a level $\alpha=0.05$.
The  joint  coordinate and block thresholding test rejects the null hypothesis because ${\boldsymbol \theta}_1$ and  ${\boldsymbol \theta}_2$ are both declared significantly different from zero.
Moreover the test provides the information that level 3 of the type of stool is not significant. 

After choosing the contrast  $\sum_{i=1}^4 \theta_{1,i}=0$, the {\tt lme} procedure available in {\tt R} also declares $\sigma_2$ significantly different from zero,
and $\theta_{1,3}$ not significant. 
\small
 \begin{table}[h!]
   \centering
   \caption{ErgoStool data: thresholded estimate at a level $\alpha=0.05$. The four fixed effects are thresholded coordinatewise and the random effects blockwise.}\label{tab:mixed}
   \begin{tabular}{l|ccccccccc} \hline
   Fixed  &$\theta_0$ & & $\theta_{1,1}$ & $\theta_{1,2}$ & $\theta_{1,3}$ & $\theta_{1,4}$ \\
   &10.2 && -0.81 & 1.38 &  0 & -0.14 \\ \hline
%   &10.2 && -0.66 & 1.23 &  0 & 0 \\ \hline
   Random  & $\theta_{2,1}$ &   $\theta_{2,2}$ &   $\theta_{2,3}$ &   $\theta_{2,4}$ &   $\theta_{2,5}$ &   $\theta_{2,6}$ &   $\theta_{2,7}$ &   $\theta_{2,8}$ &   $\theta_{2,9}$ \\
    & 0.52 &  0.52 & 0.11 &-0.30 &-0.50 & -0.03 & 0.11 &-0.57 &-0.10
%   & 0.44 & 0.44 & 0.08 & -0.27 & -0.44 & -0.03 & 0.08 & -0.50 & -0.09
    \end{tabular}
 \end{table}
 \normalsize

%\subsection{Continuous and categorical grouped predictors}
\section{Extension}
\label{sct:YuanLin}

Instead of pure testing, \citet{Yuan:Lin:mode:2006} are interested in model selection with good mean squared error performance.
To that aim, the universal threshold  of \citet{Dono94b} can be adapted to our thresholding procedure.
%where covariates are blocked and their least squares estimate correlated.
Recall that for wavelet smoothing, the matrix $X$ is orthonormal and 
the universal threshold $\lambda=\sqrt{2 \log N}$ has the property to recover the true zero-vector with high probability since
${\rm P}(\hat {\boldsymbol  \theta}_{\lambda_N}={\bf 0} \mid H_0)\approx 1-1/\sqrt{\pi \log N}$.
\citet{Dono95asym} also showed minimax results for this choice of threshold.
%where $N$ is the sample size and the number of parameters.
In the situation where $X$ is not orthonormal but is rather a complex ANOVA matrix,
one can define the quantile universal threshold.

\bigskip
{\bf Definition}: Consider model~(\ref{eq:genANOVA}) where ${\boldsymbol \theta}$ is segmented into $Q$ groups.
The quantile universal threshold (QUT) for the point estimate~(\ref{eq:gammahat}) is $\lambda_Q$ such that
\begin{equation} \label{eq:QUT}
F_{\Lambda_0}(\lambda_Q)=1-\alpha \quad {\rm with} \quad \alpha=1/\sqrt{\pi \log Q},
\end{equation}
where $F_{\Lambda_0}$ is the null distribution of the smallest threshold $\Lambda_0$ defined in (\ref{eq:pivotANOVA}) that sets to zero all estimated coefficients when the true ones are null.

\bigskip
We investigate the model selection and predictive performance of employing the quantile universal threshold (\ref{eq:QUT}) to 
threshold the parameters of the linear model (\ref{eq:genANOVA}).
We consider Models III and IV used by \citet{Yuan:Lin:mode:2006} to compare various estimators. Letting $\epsilon \sim {\rm N}(0,2^2)$, then
\begin{itemize}
 \item Model III has 2 factors out of $Q=16$: let $Z_1,\ldots,Z_{16},W$ be i.i.d.~standard Gaussian and $X_i=(Z_i+W)/\sqrt{2}$, then generate $100$ samples from
 $$
 Y=\{ X_3^3+X_3^2+X_3 \}+\{\frac{1}{3} X_6^3-X_6^2+\frac{2}{3}X_6\}+\epsilon,
 $$
 There is a total of 16 groups of size 3.
 \item Model IV has 3 factors out of $Q=20$: let $X_1,\ldots,X_{20}$ be generated as in Model~III, and let $X_{11},\ldots,X_{20}$ be trichotomized as 0, 1 and 2 if smaller then $\Phi^{-1}(1/3)$, larger than $\Phi^{-1}(2/3)$ or in between,
 then generate $100$ samples from
 $$
 Y=\{X_3^3+X_3^2+X_3\}+\{\frac{1}{3} X_6^3-X_6^2+\frac{2}{3}X_6\}+\{2 I(X_{11}=0)+I(X_{11}=1)\}+\epsilon,
 $$
  There is a total of 20 groups, half of size 3 and half of size 2.
\end{itemize}
We compare the performance of three estimators, least squares, grouped lasso and smooth blockwise iterative thresholding (\ref{eq:gammahat}) for $s=1$ (SBITE) and quantile rescaling 
based on three criteria: the number of factors selected, the MSE on the coefficients ${\boldsymbol \theta}$,
and, of marginal interest for ANOVA, the predictive MSE on ${\boldsymbol \mu}=X{\boldsymbol \theta}$.
We estimate these quantities by taking the average over 200 runs of a Monte-Carlo simulation. The results are summarized in Table~\ref{tab:modelIII}.
The selected threshold $\lambda$ for group-lasso is either $C_p$ or oracle  \citep{Yuan:Lin:mode:2006}. The SBITE estimator uses the quantile universal threshold~(\ref{eq:QUT}) instead.
The empirical results point to the excellent performance of SBITE with the quantile universal threshold both for the estimation of the number of factors and the MSE of the estimated ANOVA coefficients.
\small
 \begin{table}[h!]
   \centering
   \caption{Results of Monte-Carlo simulation of \cite[Table~1, p.~61]{Yuan:Lin:mode:2006}.
   %Columnwise, three estimators: least squares, group lasso (based on $C_p$ and oracle criteria) and SBITE with quantile universal threshold~(\ref{eq:QUT}).
   In bold, the best between $C_p$ and QUT.}\label{tab:modelIII}
   \begin{tabular}{l|ccc} \hline
           & Least squares  & Group lasso & SBITE \\
           &&($C_p$/oracle)&(QUT)\\ \hline
Model III \\
Estimated number of factors \\
out of 16. True=2. & 16                       & 11/7.5                          & {\bf 3.7}      \\
Model error \\
 \hspace{.4in} on ${\boldsymbol \theta}$ & 7.2  & 1.5/0.7          & {\bf 0.6}        \\
 \hspace{.4in} on $X{\boldsymbol \theta}$& 7.5  & 1.5/0.9          & {\bf 1.4} \\
%             & (0.2)        & (2.22)             & (0.08/0.04)                        & (0.07)      \\
%                  && (0)     & (2.11)             & (0.3/0.2)                        & (0.1) \\
\\
Model IV       \\
Estimated number of factors\\
out of 20. True=3. & 20                   & 15/10                         & {\bf 5.2}       \\
Model error \\
 \hspace{.4in} on ${\boldsymbol \theta}$ & 15                      &    3.4/2.1   &  {\bf 2.9}       \\
  \hspace{.4in} on $X{\boldsymbol \theta}$ & 5.7        & {\bf 1.6}/1.1    & 2.0 \\
%             && (2.38)        & (1.64)             & (0.92)                        & (0.89)      \\
%                  && (0)      & (2.26)             & (3.81)                        & (1.17)
   \end{tabular}
 \end{table}
\normalsize

\section{Conclusions}
\label{sct:conclusion}

Thresholding tests alleviate two problems of standard ANOVA tests: they do not require to specify types of contraint and they do not require exact knowledge of the distribution of complex pivots but simply require an estimate of the critical
value, for instance by Monte Carlo.
For the first time, block and coordinate thresholding is employed jointly to combine tests of various natures on the same parameters and therefore increase the power of the test under
different alternatives. 
Hence, observing that Fisher's test comes from block thresholding and Tukey's test comes from coordinate thresholding, we filled a possible continuum between these two tests
by developing hybrid tests based on $\ell_2$- and $\ell_\infty$-norms, essentially tests based on combined $F$- and $t$-tests. More generally $\ell_p$-tests could be derived.

How to put variables into groups is the choice of the statistician based not only on the nature of the parameters (fixed or random effects, main effects, interaction effects)
but also on the type of alternative hypothesis he or she wants to test (sparse or dense).

By deriving new tests based on thresholding in a linear ANOVA setting, this paper is the extension and practical implementation of  group lasso and the max-test. 
The level of the test can be set to any desired level, which was not addressed by the original group lasso.
We also showed that the proposed quantile rescaling is crucial to insure that parameters are democratically represented in the test whether they belong to a block of large or small size.

Combining dependent tests of various natures  could be done for the higher criticism and false discovery rate approaches as well,
 by extending the work of \cite{DonohoJin:2004} and \cite{Benj:Hoch:1995} with $p$-values related to dependent $t$-tests and $F$-tests.
 
{\tt R} code is available upon request.

% We saw that Fisher's test comes from block thresholding, and Tukey's test comes from coordinate thresholding, and that the gap can be

\section{Acknowledgements}

I thank David Donoho for an insightful discussion and for proposing the name 'O\&+ test',
Olivier Renaud for interesting discussions about ANOVA,  Caroline Giacobino for her helpful rigor, and Florian Stern \citep{Florian12} and Laura Turbatu for their contribution to an early development of the idea.

\bibliographystyle{plainnat}
\bibliography{article}

\end{document}